\documentclass[a4paper,12pt, draftcls, onecolumn]{IEEEtran}
\usepackage{graphics, graphicx}
\usepackage{amssymb,amsmath}
\usepackage{amsmath}
\usepackage[rm]{subfigure}    
\usepackage{threeparttable}   
\usepackage[dvips]{color}
\usepackage{url}   
\usepackage[utf8]{inputenc}
\usepackage{algorithm}               
\usepackage{algorithmic}    
\usepackage{cite}
\usepackage{multirow}
\usepackage[dvips]{color}
\usepackage{lineno}
\usepackage[normalem]{ulem}
\usepackage{color,soul}  

\usepackage{indentfirst}
\usepackage{setspace}
\usepackage{cases}

\newcommand{\bm}[1]{\mbox{\boldmath{$#1$}}}

\usepackage{tikz}
\definecolor{gold}{rgb}{0.85,.66,0}
\newcommand{\colb}{\textcolor{blue}}

\newcommand{\colk}{\textcolor{black}}

\begin{document}
\title{Achieving Fair Random Access Performance in Massive MIMO Crowded Machine-Type Networks}

\author{José Carlos Marinello, Taufik Abrão, Richard Demo Souza, Elisabeth de Carvalho, Petar Popovski 

\vspace{-4.5mm}

\thanks{This work was supported in part by the  Arrangement between the European Commission (ERC) and the Brazilian National Council of State Funding Agencies (CONFAP), CONFAP-ERC Agreement H2020, by the National Council for Scientific and Technological Development (CNPq) of Brazil under grants 304066/2015-0, 404079/2016-4 and 304503/2017-7, Coordenação de Aperfeiçoamento de Pessoal de Nível Superior (CAPES) under grant Programa PrInt 698503P.}
\thanks{J.C. Marinello and T. Abrão are with Electrical Engineering Department, State University of Londrina (UEL), Londrina, PR, Brazil;  \texttt{taufik@uel.br}.}
\thanks{R. D. Souza is with Department of Electrical and Electronics Engineering,  Federal University of Santa Catarina (UFSC), Florianópolis, SC, Brazil; \texttt{richard.demo@ufsc.br}.}
\thanks{E. de Carvalho and P. Popovski are with the Department of Electronic Systems,  Technical Faculty of IT and Design; Aalborg University,	Denmark; \texttt{petarp@es.aau.dk}.} 
}

\maketitle 

\begin{abstract}
The use of massive multiple-input multiple-output (MIMO) to serve a crowd of user equipments (UEs) is challenged by the deficit of pilots. Assuming that the UEs are intermittently active, this problem can be addressed by a \colk{shared access to the pilots and a suitable} random access (RA) protocol.  The strongest-user collision resolution (SUCRe) is a previously proposed RA protocol that often privileges the UEs closer to the base station (BS). \colk{In contrast,} we propose a novel RA protocol using a decentralized pilot power allocation method that aims at a fairer performance. \colk{The proposed access class barring with power control (ACBPC) protocol} allows each UE to estimate, without additional overhead, how many UEs collided for the chosen pilot and calculate an \colk{ACB} factor, which is then used to determine the pilot retransmission probability in the next protocol step. The results show that the proposed ACBPC protocol is superior to SUCRe in terms of \colk{providing a fair connectivity for very crowded networks}, although still being distributed and uncoordinated as the original SUCRe protocol. 
\end{abstract}


\vspace{-2mm}

\section{Introduction}\label{sec:intro}
\colk{In 5G networks, it is envisaged that roughly 7 billion of devices will be connected across the world by cellular technologies \cite{Ericsson2016}.} \colk{Massive multiple-input multiple-output (MIMO) techniques can be successfully used to improve the performance of random access {(RA)} in such crowded networks \cite{Bjornson.2017,Han.2017,Han.2017nov,Carvalho2017,Marinello2019}. Among them, the inventive strongest-user collision resolution (SUCRe) protocol \cite{Bjornson.2017}, by exploiting the intrinsic properties of favorable propagation in massive MIMO, offers a distributed RA solution that can resolve up to 90\% of collisions, while being able to serve a large number of users.} 

The good performance of SUCRe led to the proposal of several variants, as \cite{Han.2017, Han.2017nov, Marinello2019}. Common to them, in the first step the user equipments (UEs) transmit \colk{randomly} selected pilots. \colk{In SUCRe, only the strongest UE (the UE with the larger average channel gain) in a collision is allowed access to the select pilot and retransmits it in the third step. The other UEs in the collision wait until the next access attempt. In SUCRe combined idle pilots access (SUCR-IPA) \cite{Han.2017}, the base station (BS) detects the set of pilots that are unused in step 1 and transmits their indices with an access class barring (ACB) factor in the second step, hence creating an opportunity for the weakest UEs to access those unused pilots in the third step.}
SUCR-IPA outperforms SUCRe at the cost of increased control overhead. SUCRe combined with graph-based \colk{pilot} access (SUCR-GBPA) \cite{Han.2017nov} also announces the \colk{indices} of non-used pilots in the second step, but not an ACB factor.  Then, those UEs not apt to transmit data with the original selected pilot, randomly select another from the non-used pilots broadcasted by the BS. A bipartite graph can be constructed by setting the active UEs as the variable nodes and the chosen pilots as the factor nodes. \colk{A} successive interference cancellation  is used to estimate the channel of each UE. Through the idea of retransmission probability, a soft decision  rule for SUCRe in overcrowded scenarios is proposed in \cite{Marinello2019}. A retransmission from the $k$-th UE depends on the probability of it being the strongest contender for a pilot. \colk{The UE is able to compute this probability itself by knowing some parameters of the network, like the path-loss exponent and the number of inactive UEs in the cell. This soft SUCRe approach achieves a better RA performance\footnote{\colk{In terms of average number of access attempts and probability of failed access attempts.}} than original SUCRe protocol, without requiring additional coordination or centralized processing.} 

When a pilot collision occurs in SUCRe, the UEs that are closer to the BS, and therefore have the strongest channels, are favored in the collision resolution process. In the case of very crowded networks, this may considerably increase the probability of failed access attempts for the \colk{other} UEs, \colk{leading to an unfair performance. Similar effects can be seen with SUCRe variants, like SUCR-IPA, SUCR-GBPA, and soft SUCRe, as a consequence of selecting the \emph{strongest} UE in the collision resolution proccess. However, it is not evaluated in \cite{Bjornson.2017,Han.2017,Han.2017nov,Marinello2019} how the distance of the UE to the BS can affect the RA performance}.

The main contribution of this work is the \colk{access class barring with power control (ACBPC)} protocol, which, based on a power control (PC) policy, enables the UEs to estimate the number of contenders for a pilot, without requiring additional overhead. Based on such estimation, an ACB factor is calculated {in} a decentralized manner, which corresponds to the probability of pilot retransmission in the next step. The main advantages of the proposed ACBPC protocol are: ({\bf i}) improved \emph{fairness} in terms of connectivity, since the achieved RA performance becomes independent of the UE distance to the BS; ({\bf ii}) improved RA performance in the overcrowded scenario in comparison to SUCRe, since ACBPC has a better capability of resolving pilot collisions with many contenders; ({\bf iii}) improved spectral efficiency in comparison with conventional ACB-based protocols in which the factor is sent by the BS, \emph{e.g.}, as in \cite{Han.2017}, since in ACBPC the UEs are able to obtain the ACB factor by themselves, without additional overhead, by leveraging the channel hardening effect due to the massive number of antennas at the BS; \colk{ ({\bf iv}) improved energy efficiency  in the overcrowded scenario, since \colk{ACBPC achieves a fair RA performance with a reduced average energy consumption per UE.}} Numerical results demonstrate the effectiveness of ACBPC in very crowded scenarios, with as much as 11000 inactive UEs, since it obtains a much more uniform probability of collision resolution as a function of the distance to the BS. Besides, \colk{the ACBPC UEs consumes a lower average energy for RA purpose than UEs employing SUCRe or soft SUCRe protocol at any distance to the BS, while this energy saving can reach $88\%$ at certain distances.} 
 
\vspace{-2mm}
\section{System Model}\label{sec:syst}
\vspace{-.2mm}

Assume a \colk{hexagonal} cellular network in which the BSs operate in time-division duplexing and have $M$ antennas. Time-frequency resources are divided into coherence blocks of $T$ channel uses\colk{; most blocks are used for payload data transmission for active UEs, which have been allocated dedicated pilots, while a few blocks are dedicated for RA \cite[Fig. 2]{Bjornson.2017}.}  $\mathcal{U}_i$ is the set of UEs in cell $i$, while $\mathcal{A}_i \subset \mathcal{U}_i$ is the subset of {UEs} active in this cell at a given time. In very crowded machine-type communications scenarios,  $|\mathcal{U}_i|\gg T$, but UEs become active with probability $P_a \leq 1$. Thus, $|\mathcal{A}_i|< T$  \colk{holds}, and therefore the BS can temporarily assign orthogonal pilots to the active UEs. By focusing on cell 0, and denoting $\mathcal{K}_0 = \mathcal{U}_0 \backslash \mathcal{A}_0$ as the set of inactive UEs, \colk{its cardinality} is $K_0 = |\mathcal{K}_0|$. Such \colk{UEs} share $\tau_p$ orthogonal \colk{RA} pilots $\bm \psi_1\ldots \bm \psi_{\tau_p} \in \mathbb{C}^{\tau_p}$ 
satisfying $||\bm \psi_t||^2 = \tau_p$, $t \in \{1,2,\ldots,\tau_p\}$. 
The UEs \colk{wanting to} become active randomly choose one of the available $\tau_p$ \colk{RA} pilots in each RA attempt; {\it e.g.}  the $k$th UE selects pilot $t$ \colk{and transmits it with power $\rho_k > 0$}.  Therefore, if the set \colk{$\mathcal{S}_t \subset \mathcal{K}_0$} contains the \colk{indices} of the UEs that selected \colk{RA} pilot $t$, then $|\mathcal{S}_t| \sim {\mathfrak{B}}\left( K_0, \frac{P_a}{\tau_p} \right)$ \cite{Bjornson.2017}. 

The channel between the $k$-th UE and  BS 0 is ${\bf h}_k = \sqrt{\beta_k} \underline{{\bf h}}_k  \in \mathbb{C}^M$, where $\beta_k$ represents large-scale fading, consisting of path loss and shadowing, while small-scale fading $\underline{{\bf h}}_k \sim \mathcal{CN}(0,\colb{{\bf I}_M})$. Moreover, it is reasonable to assume that large-scale fading $\beta_k$ is known, once it is possible to estimate it, for instance, in the \colk{SUCRe protocol initialization step} \cite{Bjornson.2017}. 

Focusing on \colk{the RA stage}, \colk{the four steps of SUCRe are:}

\noindent\colk{\underline{Step \bf 1}:} the BS receives ${\bf Y} \in \mathbb{C}^{M \times \tau_p}$
\begin{equation}\label{eq:Rx_pilot}
{\bf Y} = \colk{\sum_{t=1}^{\tau_p}} \sum_{k \in \mathcal{S}_t} \sqrt{\rho_k} {\bf h}_k {\bm \psi}^T_{t} + {\bf W} + {\bf N}, 
\end{equation}
where ${\bf N} \in \mathbb{C}^{M \times \tau_p}$ is noise with each element distributed as $\mathcal{CN}(0,\sigma^2)$, while ${\bf W} \in \mathbb{C}^{M \times \tau_p}$ is the intercell interference. Then, the BS correlates ${\bf Y}$ with $\bm \psi_t$ to obtain
\begin{eqnarray}\label{eq:csi_est}
{\bf y}_t &=& {\bf Y} \frac{\bm \psi_t^*}{||\bm \psi_t||} = \sum_{i \in \mathcal{S}_t} \sqrt{\rho_i \tau_p} {\bf h}_i + {\bf W}\frac{\bm \psi_t^*}{||\bm \psi_t||} + {\bf n}_t,
\end{eqnarray}\colk{where the effective noise is ${\bf n}_t = {\bf N}\frac{\bm \psi_t^*}{||\bm \psi_t||} \sim \mathcal{CN}({\bf 0}, \sigma^2 {\bf I}_M)$.} 
\noindent\colk{\underline{Step \bf 2}:} BS transmits with power $q$, orthogonal precoded {downlink (DL)} signal  ${\bf V} \in \mathbb{C}^{M \times \tau_p}$ according to each pilot:
\begin{equation}\label{eq:DL_signal}
{\bf V} = \sqrt{q} \sum_{t=1}^{\tau_p} \frac{{\bf y}_t^*}{||{\bf y}_t||} {\bm \psi}_t^T.
\end{equation}
The UEs receive ${\bf z}_k \in \mathbb{C}^{\tau_p}$, $k \in \mathcal{S}_t$
\begin{equation}\label{eq:DL_rec}
{\bf z}_k^T = {\bf h}_k^T {\bf V} + {\bm \nu}^T_k + {\bm \eta}^T_k,
\end{equation}
${\bm \nu}_k \in \mathbb{C}^{\tau_p}$ is inter-cell interference, and ${\bm \eta}_k \sim \mathcal{CN}({\bf 0}, \sigma^2 {\bf I}_{\tau_p})$ is noise. After correlating ${\bf z}_k$ with ${\bm \psi}_t$, the UE calculates
\begin{equation}\label{eq:DL_rec_corr}
z_k = {\bf z}_k^T \frac{\bm \psi_t^*}{||\bm \psi_t||} = \sqrt{q \tau_p}{\bf h}_k^T \frac{{\bf y}_t^*}{||{\bf y}_t||} + {\bm \nu}^T_k \frac{\bm \psi_t^*}{||\bm \psi_t||} + \eta_k,
\end{equation}
where $\eta_k \sim \mathcal{CN}(0, \sigma^2)$. 
Let $\alpha_t = \sum_{i \in \mathcal{S}_t} \rho_i \beta_i \tau_p + \omega_t$  be the sum of gains and interference as seen at the BS according to  \eqref{eq:csi_est}, then an asymptotically error free estimator for $\alpha_t$ is 
\cite{Bjornson.2017}:
\begin{equation}\label{eq:alpha_est}
\hspace{-.4mm}\hat{\alpha}_{t,k} = \max\left(\left[\frac{\Gamma(M+\frac{1}{2})}{\Gamma(M)}\right]^2 \frac{q \rho_k \beta^2_k \tau^2_p}{[\Re(z_k)]^2} - \sigma^2, \,\,\rho_k \beta_k \tau_p\right), \hspace{-2mm}
\end{equation}
$\Re(\cdot)$ is the real part and $\Gamma(\cdot)$ is the complete Gamma function.

\noindent\colk{\underline{Step \bf 3}:} SUCRe resolves pilot collisions in a distributed fashion, such that each pilot is retransmitted, ideally, by only one UE. Notice that each UE $k \in \mathcal{S}_t$ knows its own average signal gain $\rho_k \beta_k \tau_p$ and \colk{obtains} an estimate $\hat{\alpha}_{t,k}$ for \colk{$\alpha_t$ evaluating \eqref{eq:alpha_est}}. Then, each UE applies locally the \emph{decision rule}: 
%
\begin{eqnarray}
\begin{aligned} \label{eq:sucre_hard_dec}
\mathcal{R}_k: \quad {\text{if}} \quad & \rho_k\beta_k\tau_p  >  \frac{\hat{\alpha}_{t,k}}{2} + \epsilon_k \qquad \text{(repeat)}\\
\mathcal{I}_k: \quad {\text{if}} \quad  & \rho_k\beta_k\tau_p  \leq  \frac{\hat{\alpha}_{t,k}}{2} + \epsilon_k \qquad \text{(inactive).}  
\end{aligned}
\end{eqnarray}
\colk{A suitable value for  the bias parameter $\epsilon_k \in \mathbb{R}$ in \eqref{eq:sucre_hard_dec} is given in \cite{Bjornson.2017} based on $\overline{\omega}$, which is the average value of $\omega_t$.}

\noindent\colk{\underline{Step \bf 4}: The BS allocates dedicated data pilots to the UEs if no RA collision remained.}

\section{Proposed RA Protocol}\label{sec:RAPPPC}
A drawback of SUCRe is that it gives preference to the strongest UE when resolving a collision, which usually results in selecting those closer to the BS, in detriment of the edge UEs. As the number of inactive UEs grows, and collisions occur more often, it becomes difficult to the edge UEs {to} establish a connection with the network. Instead, in this work we propose \colk{an RA protocol that introduces fairness among UEs}, in the sense that the UEs have basically the same success connection probability, \colk{regardless of their distances to the BS}.

Since the $k$th UE has an estimate of its long-term fading coefficient, $\beta_k$, it can {adjust} its pilot transmit power inversely proportional to $\beta_k$, \colk{up to a maximum transmit power limit}. Therefore, the pilot transmit power of the $k$th UE is \colk{$\rho^{\rm pc}_k = \min \{\frac{\overline{\rho}}{\beta_k}, \rho^{\rm max}\}$}, 
%
%
in which $\overline{\rho}$ is the average received power at the BS, \colk{and $\rho^{\rm max}$ is a maximum transmit power constraint}. \colk{The signal received at the BS and its correlation with $\bm\psi_t$ results \colk{in expressions} similar to \eqref{eq:Rx_pilot} and \eqref{eq:csi_est}, respectively, but considering $\rho_k^{\rm pc}$ instead of the uniform power coefficients $\rho_k$ \colk{adopted in SUCRe}. Indeed, the power control policy implies that the BS estimates only small-scale fading coefficients\colb{\footnote{When $\frac{\overline{\rho}}{\beta_{k}} > \rho^{\rm max}$ for a given UE, the small-scale fading vector of this UE in ${\bf y}_t$ under our power control policy will be weighted by a factor lower than one, but the protocol proceeds in the same way. Since such condition occurs very rarely, it does not result in any significant performance loss.}} for the UEs in $\mathcal{S}_t$ when evaluating \eqref{eq:csi_est}.}
%
%
%
In the second step, the precoded DL pilot signal sent by the BS is computed as in \eqref{eq:DL_signal}; then, the received signal at UE $k$ is according to \eqref{eq:DL_rec}, upon which $z_k$ is obtained as in \eqref{eq:DL_rec_corr}. Finally, the {\it sum of gains and interference} at the BS is
\begin{eqnarray}\label{eq:alpha2}
\alpha_t = \sum_{i \in \mathcal{S}_t} \rho^{\rm pc}_i \beta_i \tau_p + \omega_t = \overline{\rho} \tau_p |\mathcal{S}_t| + \omega_t,
\end{eqnarray}
and is estimated at the UEs as \eqref{eq:alpha_est} \colk{with $\rho^{\rm pc}_k$ replacing $\rho_k$}.

The interference $\omega_t$ can be partially canceled by subtracting its average $\overline{\omega}$ from $\hat{\alpha}_{t,k}$. Notice that $\overline{\omega}$ can be estimated since it has the same fixed value for all UEs \cite{Bjornson.2017}. We can thus obtain an estimate for $|\mathcal{S}_t|$ from $\hat{\alpha}_{t,k}$ under the proposed approach as
\begin{equation}\label{eq:St_est}
\widehat{|\mathcal{S}_t|}_k = \frac{\hat{\alpha}_{t,k} - \overline{\omega}}{\overline{\rho} \tau_p}.
\end{equation}

Thereby, an ACB factor can be calculated as $\colk{\zeta}_k ={\widehat{|\mathcal{S}_t|}_k}^{-1}$, so that each UE retransmits its pilot in step 3 with probability $\colk{\zeta}_k$. \colk{Note that the proposed solution does not require any additional control overhead for the UEs obtaining the ACB factor}, in contrast to other ACB-based RA protocols \cite{Han.2017}. \colk{Besides, ACBPC demands little additional computation compared to SUCRe. It is only required \colk{that the UEs take} a decision with probability $\colk{\zeta}_k$, which can be done by generating a random number $\in[0,1]$ and comparing it with $\colk{\zeta}_k$.}

It is simple to derive an upper bound for the probability of ACBPC protocol resolving a pilot collision, $P_{\text{res}}$, \colk{with $|\mathcal{S}_t|$ contenders. The obtained expression constitutes an upper bound since it assumes a perfect estimation for $|\mathcal{S}_t|$}. The probability of pilot retransmission in step 3 of SUCRe is $\frac{1}{|\mathcal{S}_t|}$, and the collision is resolved if only one UE retransmits the $t$th pilot. Therefore, we have
\begin{eqnarray}\label{eq:P_resolved}
P_{\text{res}} = {{|\mathcal{S}_t|}\choose{1}} \frac{1}{|\mathcal{S}_t|} \left(1 - \frac{1}{|\mathcal{S}_t|} \right)^{|\mathcal{S}_t| - 1} = \left(1 - \frac{1}{|\mathcal{S}_t|} \right)^{|\mathcal{S}_t| - 1}. \hspace{-4mm}
\end{eqnarray}Given the representation of the exponential function as a limit, $e^x = \lim_{n \to \infty} \left( 1 + \frac{x}{n} \right)^n$, then $\lim_{|\mathcal{S}_t| \to \infty} P_{\text{res}} = e^{-1}$. This implies that the probability of \colk{ACBPC} resolving pilot collisions with so many contenders converges to $36.78 \%$, differently than SUCRe in which this probability tends to 0.

\section{Numerical Results and Discussion}\label{sec:results}


If a collision occurs and the protocol is not able to resolve it, the UE makes a new attempt after a random backoff \colk{as in \cite{Bjornson.2017}}. In our setup, if the number of access attempts \colk{from} a UE exceeds 10, a failure is declared. We compare the performance of the proposed ACBPC protocol with that of SUCRe \cite{Bjornson.2017} \colk{and/or soft SUCRe \cite{Marinello2019}}, considering \colk{different performance metrics}. \colk{Our focus on \colk{these protocols}, instead of SUCR-IPA and SUCR-GBPA, is because \colk{the latter ones} require significant additional overhead for the BS transmitting the idle pilots \colk{indices} to the UEs. This overhead is harmful to the system performance from both spectral and energy viewpoints. In contrast, SUCRe, \colk{soft SUCRe,} and the proposed ACBPC protocol require the same level of knowledge at the UE, allowing a fair comparison.} Moreover, $P_a = 0.1\%$, $\tau_p = 10$, and $K_0 \in [1,\colk{28000}]$. The other parameters are as in \cite{Bjornson.2017}: $d_{\text{max}}=250$m, $d_{\text{min}}= 25$m, $\overline{d} = 10^{-3.53}$, path loss exponent $\kappa =3.8$, shadowing standard deviation $\sigma_{\rm shadow} = 8$ dB\colk{, and $\beta_k = \overline{d} \cdot d^{-\kappa} \cdot \chi_k$, with $d_k$ being the distance and $\chi_k$ the log-normal shadowing coefficient of the $k$th UE.} \colk{The intercell interference is obtained considering $\tau_p$ active UEs randomly positioned in each one of the 6 adjacent cells, and transmitting with the same average power than UEs in cell 0.} \colk{In the SUCRe protocol, the transmit power is fixed and computed in the same way as in \cite{Bjornson.2017}, \emph{i.e.}, $\rho_k = \rho^{\rm SUCRe} = \sigma^2/(\overline{d} \cdot d_{\text{max}}^{-\kappa})$, $\forall k$. On the other hand, for the proposed ACBPC protocol, we have set $\overline{\rho} = \sigma^2$, and $\rho^{\rm max} = \rho^{\rm SUCRe}$, assuring that ACBPC UEs never transmit with more power than SUCRe UEs.}

\begin{figure}[!htbp]
\centering
\includegraphics[width=.795\textwidth]{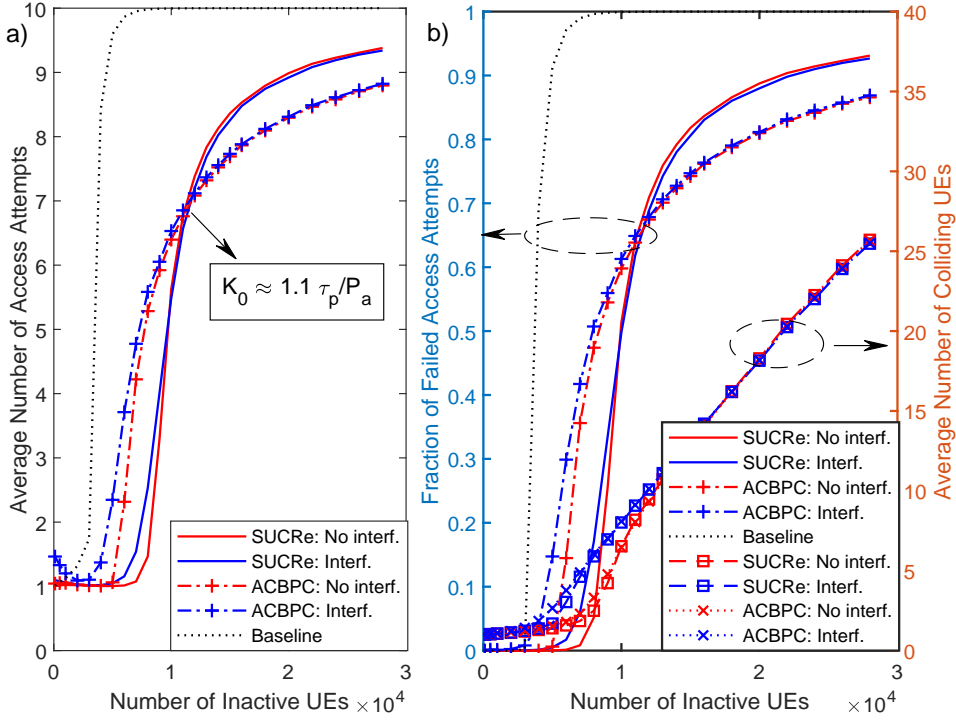} 
\vspace{-4mm}
\caption{Average number of access attempts (a) and probability of failed access attempts \colk{and average number of colliding UEs} (b) in a crowded network as a function of the number of devices for the ACBPC and SUCRe protocols. \colk{The baseline scheme is a conventional protocol where pilot collisions are only handled by retransmission in later RA blocks \cite{Bjornson.2017}.}}
\label{fig:Perf_SUCRe}
\end{figure}

Fig. \ref{fig:Perf_SUCRe} shows the average number of access attempts and the probability of failed attempts for the proposed ACBPC protocol and SUCRe. \colk{Intercell interference limits the system performance for low values of $K_0$, but is not the most stringent factor in very crowded scenarios.} One can note that ACBPC is outperformed by SUCRe for $K_0 < 11000$, while ACBPC presents a better performance for overcrowded scenarios where $K_0 > 11000$. Note that when $K_0 > \tau_p/P_a$, there are on average more UEs trying to access the network than available pilots, and thus collisions \colk{are highly likely to} occur. For example, while the UEs require on average 8.275 access attempts employing SUCRe (considering interference) for $K_0 = 15000$, this number is reduced to 7.731 attempts under ACBPC protocol. The fraction of failed access attempts is reduced from 0.8082 for SUCRe to 0.7468 for ACBPC in the same conditions, which results in $0.0614 \times K_0 \times P_a = 0.921$ additional connected UEs on average when employing the proposed protocol. \colk{The figure also depicts the average number of UEs contending for an RA pilot, which shows an almost linear growth with $K_0$.}


\noindent{\it Remark 1:} One can see from Fig. \ref{fig:Perf_SUCRe} that the proposed ACBPC protocol outperforms SUCRe for $K_0 > 1.1 \frac{\tau_p}{P_a}$, while the opposite holds under this threshold. In practice, this makes \colk{it} convenient to implement the ACBPC protocol only for very crowded networks. Equivalently, one can design a switching procedure between ACBPC and SUCRe protocols depending on the number of inactive UEs, if this information is available at the BS. Indeed, $K_0$ changes very slowly with time, and can be estimated in different ways. If employing the ACBPC protocol, the BS can easily evaluate the average number of contending UEs, based on \eqref{eq:alpha2}. As the expected value of contending UEs depends on $K_0$, an estimator can be conceived based on the method of moments. If employing SUCRe, the BS can evaluate the average number of idle pilots, which is also dependent on $K_0$, allowing thus its estimation. In both cases, as $K_0$ changes very slowly with time, the estimation can be evaluated based on a very large number of observations.

\begin{figure}[!htbp]
\centering
\hspace{-2mm}\includegraphics[width=.795\textwidth]{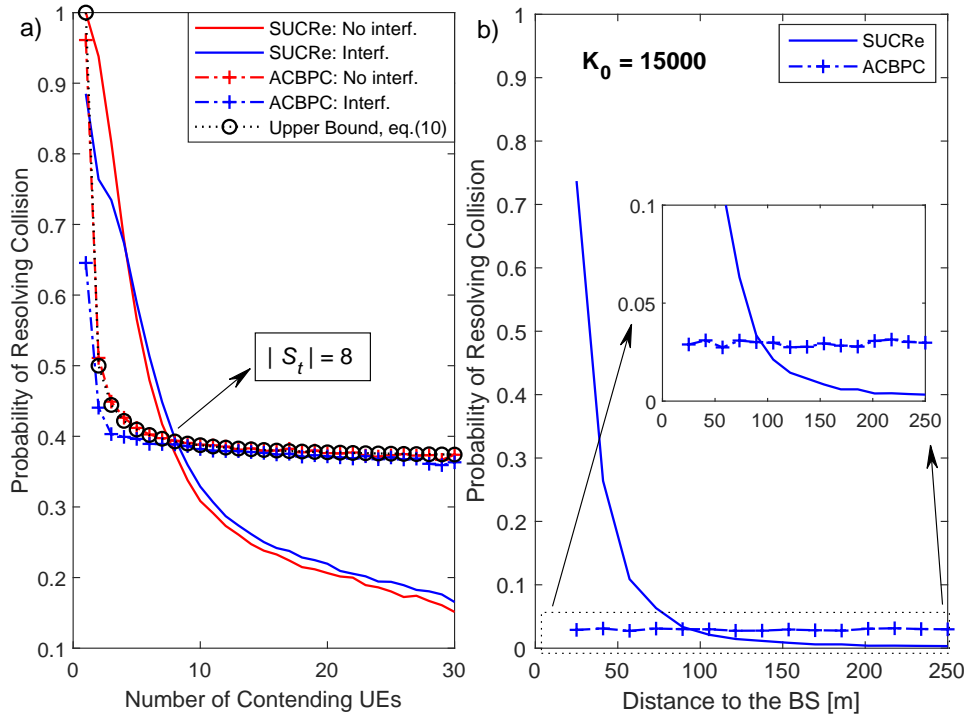}
\vspace{-5mm}
\caption{Probability of resolving collisions as a function of: \colk{a) $|\mathcal{S}_t|$; b) distances to the BS with $K_0 = 15000$.}}
\label{fig:Perf_St}
\end{figure}

Fig. \ref{fig:Perf_St}.a depicts the probability of the protocols resolving pilot collisions as a function of $|\mathcal{S}_t|$. It can be seen that SUCRe is more effective for resolving collisions with $|\mathcal{S}_t| \leq 7$, while the proposed ACBPC protocol has a greater capability of resolving collisions with $|\mathcal{S}_t| > 8$. This justifies the performance results in Fig. \ref{fig:Perf_SUCRe}, since for low values of $K_0$, the collisions occur more often with a number of contenders that SUCRe is more efficient to resolve. The opposite holds for high values of $K_0$. Note that the results  in Fig. \ref{fig:Perf_St}.a \colk{are independent of} $K_0$, which just determines what values of $|\mathcal{S}_t|$ are more common in the pilot collisions. \colk{Fig. \ref{fig:Perf_St}.a} also depicts the upper bound in \eqref{eq:P_resolved}, which almost equals the performance in the scenario without interference. Therefore, intercell interference \colk{has} the most adverse effect for the $|\mathcal{S}_t|$ estimation. \colk{For example, with $|\mathcal{S}_t| = 1$, the intercell interference can be seen by the BS as a second UE contending by that RA pilot, resulting in performance loss even in this case.} Notice that due to the channel hardening  
$\frac{||{\bf y}_t||^2}{M} \stackrel{M \to \infty}{\longrightarrow} \alpha_t + \sigma^2,
$ but $\omega_t$ does not converge to $\overline{\omega}$, since the last is averaged with respect to the adjacent cells UE locations and shadow fading realizations. Therefore, subtracting $\overline{\omega}$ from $\hat{\alpha}_{t,k}$ will not eliminate the intercell interference even with infinite number of BS antennas. Besides, further elaborated estimators for $|\mathcal{S}_t|$ can be conceived, but this is outside the scope of this paper.

On the other hand, Fig. \ref{fig:Perf_St}.b depicts the probability of a given UE winning a pilot collision as a function of its distance to the BS, with $K_0 = 15000$. \colk{Since in this case the scenarios with and without interference presented very similar results for each protocol, we omitted the latter.} As expected, SUCRe presents a very unfair behavior in the sense that the UEs closer to the BS have higher collision resolution probabilities, in detriment of the edge UEs. \colk{The} proposed ACBPC protocol presents a more uniform performance, although the collision resolution probabilities of the UEs \colk{closer than $\approx 90$m of the BS} become lower than that for SUCRe protocol. For $K_0 = 15000$, there are on average \colk{2196 UEs in the range of $[25, 90]$m, and 12804 UEs ($\approx 85.4\%$) in the range of $[90, 250]$m.} This implies that much more UEs are \colk{given access} when employing ACBPC instead of SUCRe. 

\begin{figure}[!htbp]
\centering
\includegraphics[width=.795\textwidth]{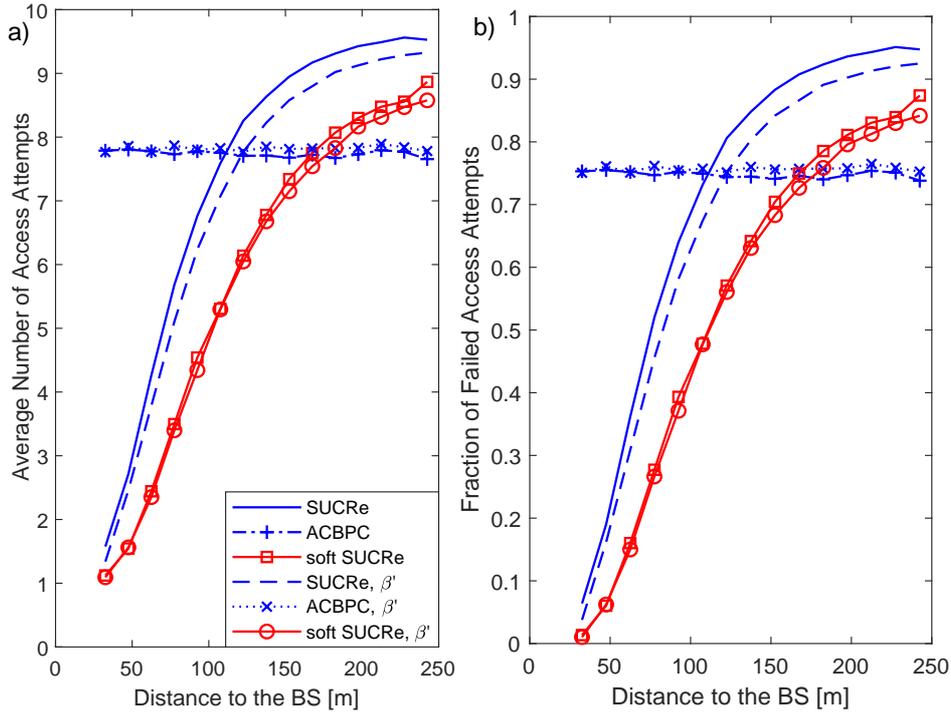} 
\vspace{-4mm}
\caption{\colk{Average number of access attempts (a) and probability of failed access attempts (b) in a crowded network as a function of the distance from the UE to the BS with \colk{interference and} $K_0 = 15000$.}}
\label{fig:Perf_SUCReDist}
\end{figure}

Since the probability of winning a pilot collision decreases for the farther UEs under the SUCRe protocol in the overcrowded scenario, one can expect an increased average number of access attempts for these UEs, as well as an increased probability of failed attempts. This is shown in Fig. \ref{fig:Perf_SUCReDist} \colk{only for the scenario with interference, since the performance curves without interference were very similar in this case}. One can see that the very unfair performance of SUCRe can be replaced by a much more uniform experience when employing the proposed ACBPC protocol. An average number of about 7.73 access attempts can be assured for the UEs in any location within the cell employing ACBPC, as well as a constant probability of failed access attempt of about 0.75. In contrast, SUCRe provides a better performance for the UEs closer than \colk{$\approx$100m} from the BS, but a very worse performance for the remaining UEs, \colk{corresponding to $\approx 80\%$ of the cell area}. Fig. \ref{fig:Perf_SUCReDist} also investigates the RA performance 
when UEs have imperfect $\beta$ estimates modeled as $\beta'_k = \varphi \beta_k$, in which $\varphi \sim \mathcal{N}(1,\sigma^2_{\beta})$, with $\sigma_{\beta} = 0.2$. It is shown that the performance of ACBPC is little sensitive to imperfect $\beta$ estimates, while SUCRe performance is slightly improved, in a similar effect than the random powers discussed in \cite{Bjornson.2017}. \colk{Fig. \ref{fig:Perf_SUCReDist} also depicts the soft SUCRe performance, \emph{i.e.}, the SUCRe with soft retransmission criterion of \cite{Marinello2019}. This latter provides a better performance for closer UEs in detriment of the farther ones, similarly to SUCRe. The distance, however, in which the UEs employing this protocol start to require a higher average number of access attempts than ACBPC UEs is increased to $\approx$ 160m, such that $\approx 49 \%$ of the UEs are closer than this distance.}

\begin{figure}[!htbp]
\centering
\includegraphics[width=.795\textwidth]{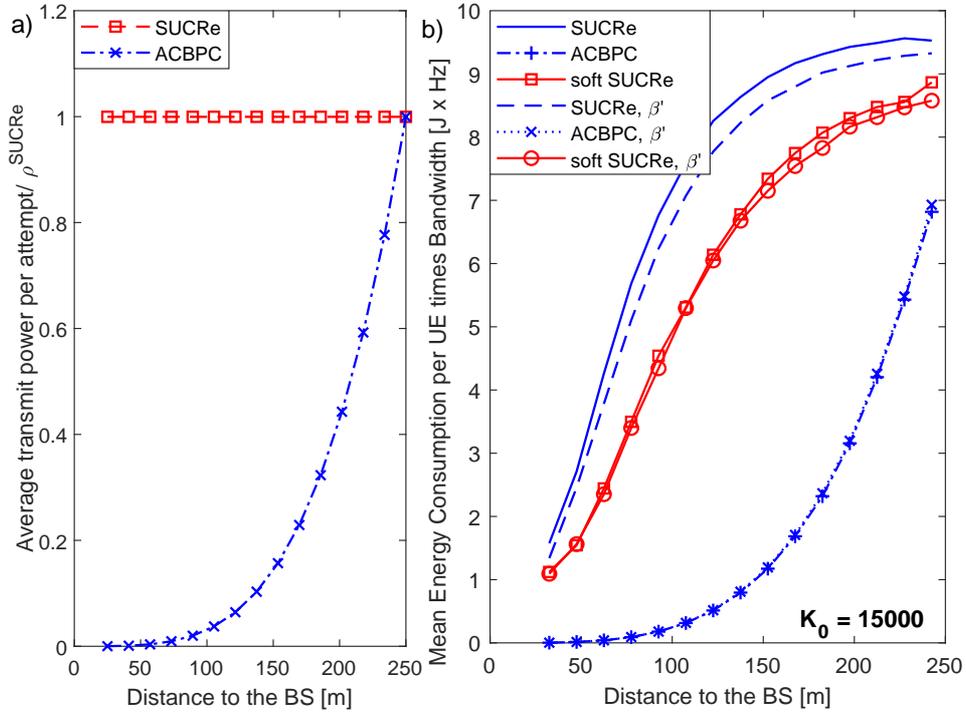} 
\vspace{-4mm}
\caption{\colk{Average transmit power per attempt (normalized by $\rho^{\rm SUCRe}$) (a), and average energy consumption times bandwidth (b) in a crowded network as a function of the distance to the BS with interference and $K_0 = 15000$.}}
\label{fig:AvEnergy}
\end{figure}

\colk{Finally, Figure \ref{fig:AvEnergy}.a depicts the average transmit power of the UEs \colk{per attempt} according to their distances to the BS, normalized by $\rho^{\rm SUCRe}$. It shows that a much lower power is transmitted on average with the power control strategy deployed by ACBPC. The soft SUCRe protocol of \cite{Marinello2019} employs the same transmit power as the SUCRe protocol, and, therefore, it is higher than the average ACBPC transmit power. On the other hand, Figure \ref{fig:AvEnergy}.b shows the average transmit energy times bandwidth\footnote{\colk{In order to obtain the final energy consumption, one should divide the depicted values by the bandwidth allocated for RA.}} per UE according to their distance to the BS, considering $\rho^{\rm max} = \rho^{\rm SUCRe} = 0.1$ W and $\tau_p = 10$. Note that the average transmit energy accounts for both the average transmit power per attempt of Fig. \ref{fig:AvEnergy}.a and the average number of access attempts of Fig. \ref{fig:Perf_SUCReDist}.a. It is shown that ACBPC UEs transmit less energy on average than SUCRe and soft SUCRe UEs, for all the distance range within the cell. For example, the UEs at the distance of 137.5m to the BS employing ACBPC spend $12\%$ of the energy they would spend employing soft SUCRe. This contributes to a better energy efficiency of the devices when employing ACBPC, as well as an uniform connectivity performance regardless of their distance to the BS.}

\section{Conclusion}\label{sec:concl}
The recently proposed SUCRe constitutes a very efficient RA protocol for massive MIMO crowded networks, but it is widely known to provide a very unfair performance for the UEs depending on their distances to the BS, and a limited performance in the overcrowded scenario. The ACBPC protocol proposed in this \colk{letter} assures a uniform RA performance for the UEs within the cell, independently of their distances to the BS, even with imperfect $\beta$ estimates. In addition, an improved RA performance is achieved in the overcrowded scenario in comparison to SUCRe, due to the better ability of our proposed approach in resolving pilot collisions with a high number of contenders \colk{spending less energy, thanks to its higher connectivity efficiency}.
The proposed ACBPC scheme is promising and can be combined with SUCRe in a hybrid solution to cover a wide parameter range for crowded scenarios.

\end{document}